# Applying Vector Space Model (VSM) Techniques in Information Retrieval for Arabic Language

Bilal Ahmad Abu-Salih

# Abstract

Information Retrieval (IR) allows the storage, management, processing and retrieval of information, documents, websites, etc. Building an IR system for any language is imperative. This is evident through the massive conducted efforts to build IR systems using any of its models that are valid for certain languages. This report presents an implementation for a core IR technique which is Vector Space Model (VSM). We have chosen VSM model for our project since it is a term weighting scheme, and the retrieved documents could be sorted according to their relevancy degree. One other significant feature for such technique is the ability to get a relevance feedback from the users of the system; users can judge whether the retrieved document is relative to their need or not. The developed system has been validated through building an Arabic IR website using server side scripting. The experiments verifies the effectiveness of our system to apply all techniques of vector space model and valid over Arabic language.

# Contents

# List of Figures

# List of Tables

# Chapter 1

# Introduction

## 1.1 Information Retrieval - an overview

Information retrieval (IR), one of the NLP advanced techniques, is defined to be the science of enhancing the effectiveness of term-based document retrieval. [44]

It could be also defined as " finding material (usually documents) of an unstructured nature (usually text) that satisfies an information need from within large collections (usually stored on computers)." [12]

Information retrieval systems (IRS) are frequently engineered, optimized and implemented mainly for English language. However, every Language has some special or common features which could be covered by information retrieval techniques with some enhancement. [10] The IR and related methods -will be discussed later-have been used mainly to search for information within documents stored in certain repository of documents (text collection).

### 1.1.1 Information versus Data Retrieval

Data Retrieval is the process of determining which documents match the user query which is not enough to satisfy the user specific need. In deed, the user in the information retrieval systems is retrieved information that is concerned with the subject of the user rather than information that only retrieved by specific query .

The data retrieval deals with data that has well defined structure and semantics. But in information retrieval, it deals with natural language text. [39] We could conclude by saying that the main objective of IR system is to retrieve all the documents which are relevant to a user query while retrieving as few non-relevant documents as possible. [17]

### 1.1.2   User Task

The user who is looking for some information in a text collection just has to translate his need to some query and pass it to the information retrieval system, this process implies specifying a set of words which convey the semantics of the information need. [39]

In addition, the user of the IR system could look for some document or set of documents by using the browsing mechanism that could be offered by the IR systems. Figure 4.1 shows the interaction between the user and the IR system.

A document is described to be relevant if the user makes a decision that the retrieved information gives him an added value with respect to his personal information need. [12]

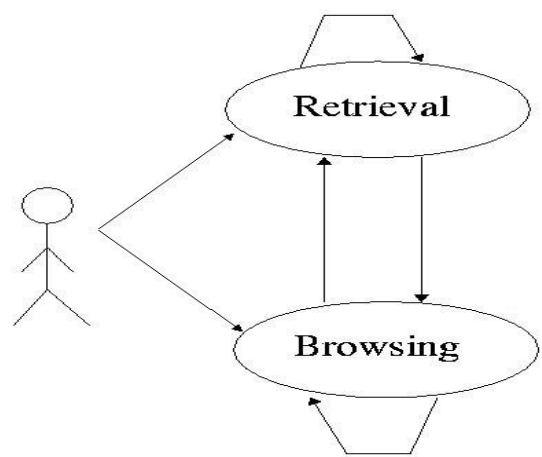

Figure 1.1: Interaction of the user with the IR system

### 1.1.3 The Retrieval Process

The process that starts with the user need and ends up with retrieving documents is considered to be the core of the IR system. Figure 4.2 illustrates the main phases of the retrieval process.

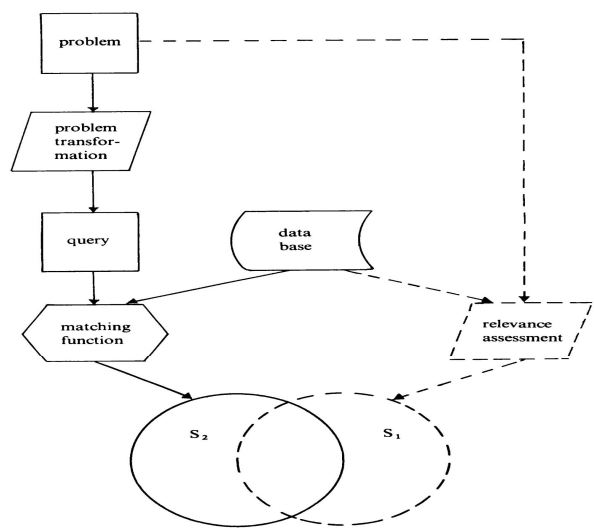

Figure 1.2: IR system Process

before the retrieval process can even be initiated, it is important to define the database. that is usually done by the manager of the database, which is responsible about doing the following: (a) the documents to be used, (b) the operations to be performed on.the text, and (c) the text model. [39]

As shown in the figure above, the IR process starts with the specifying the problem (or the user need), this will be transformed to some query. The system will search in the database for such query and get back the result to the user who will specify the relevancy of the document to his/her need. We will explain this process in details in the implementation and design chapter.

Query Operations

Queries are formal statements of user need which is put to the IR system to look for some information. The operations on queries are a function that is done by the IR system. One common query operation is called parsing, in which the query is sliced into constituent elements to be passed to the IR system. [48]

Although the user query is an element that determines an information need, but it is not the only one. There are many researches to find other methodologies, such as queryspecific contexts, to work around the query in order to find another form to represent the query in away it gives much relevant document. [24]

Tokenization and Term Operations

Tokenization is a primary step in dealing out with textual data prior the tasks of information retrieval, text mining, and natural language processing NLP. Tokenization is a language-dependent approach, including normalization, stop list removal, and stemming. [9]

The main term operations in an IR system include stemming, truncation, weighting , stoplist removal and thesaurus operations. [48] stemming is the process of making a match between the user query and document terms which are related to the same meaning but which appear in different morphological forms. [33]. For example, a searcher interested in finding documents about science may enter the word "Scientist" which would match terms such as science, scientific, and scientist and so on.

The term weighting methods is the discriminating measure that is used to assign appropriate weights to the terms which could improve the performance of text retrieval performance as well as the text categorization [43] which is "the classification of text documents into a set of one or more categories" [28]. This process will be discussed in the implementation section later on.

Index File

One of the main phases of the information retrieval process is the indexing phase, in which every document uploaded to the information retrieval system is sliced to the main words of the document (called the index terms), the file that includes these terms is called inverted file (or index file). This process is the most critical techniques, it consists of choosing the main terms to represent the documents The inverted file is created to handle a quick access to retrieve documents by using index terms, and this process could reduce the memory size. [17] There are three Kinds of indexing: 1. Word indexing 2. Stem indexing and 3. Root indexing.

In word indexing, before starting any indexing process, we must take care that there are no spelling or syntax errors in the documents by checking them one by one and correct the spelling.

In stem indexing,In a navest approach, we could treat a word as a term. Yet, morphological variants like edema and edemas are so closely related that they are usually conflated into a single word stem, e.g., edem, by stemming. [13]

By using stemming, the relevancy of the retrieved documents will be rectified and their number will also be enlarged. [17] In root indexing, we could define the root to be the pure word after removing all the suffixes , prefixes and infixes which could be used to improve the performance of the search by extracting the roots of the words in the query and match them with the index terms which have been already converted to roots.

Document Operations

Documents are essential objects in IR systems and their numerous operations are considered crucial in information retrieval process . In many types of IR systems, documents added to a database must be given distinctive identifiers, parsed into their component fields, and those fields sliced into field identifiers and index terms. [48]

One of the main document operations is the documents classification; which means classifying the documents in the text collection based on their subjects or scientific fields; manual classification could be used to classify documents in text collection such as machine learning-based text classification. [12] Another significant document operation is presenting the documents. The user interface of an IR system, as with any other type of systems, is serious to its successful handling. [48]

### 1.1.4    Models of information retrieval

IR systems are categorized into three main categories : (1). Boolean model: " based on Boolean Logic and classical Set theory in that both the documents to be searched and the user's query are conceived as sets of terms." [1] (2). Probabilistic model in which the relevance of a document for a given query could be estimated by using the probability of finding relevant information and the probability of finding a non relevant information. [31]

The process of probability estimation describes how probabilities should be predictable of the occurrence of terms in the real text collection. [21]

(3). Vector space model: in this model the document and the query both are represented in vector spaces, and this model will be the main core of our study.

Boolean model

The Boolean retrieval model is a form for information retrieval in which we can create any query that in a Boolean expression terms structure, that is, in which terms are combined with the operators AND, OR, and NOT. The Boolean model views each document as just a set of words. [12]

Boolean expressions are created and produced from queries. They represent a demand to determine what documents contain (or do not contain) a given array of keywords. For example:Find all documents include "teacher".

Reference to the definition of the Boolean model, a query searches a set of documents to verify their content. What we have already mentioned about boolean expression is therefore usually represented as the Boolean expression. Some queries attempt to find documents that do not contain a specific term or pattern. This is done by using the "not" operator. For example, the query: Find all documents that do not contain "teacher". Most queries involve looking for more than one term. For example, a user might say any of the following: Find all documents containing " teacher" and "student".

Find all documents containing " teacher " or " student " (or both).

Find all documents containing " teacher "or " student ", but not both.

The first query will retrieve only documents containing both "teacher" and "student", whereas the second will be satisfied by a document that includes one of the two words. The third query could be satisfied only by the documents that represent the union of the third and first sets of documents. [48]

Vector Space Method

In vector-space model, a document is theoretically represented by a vector of index terms exported from the text, with related weights which represent the importance of the index terms in the document and within the whole document collection; likewise, a query is modelled as a list of index terms with related weights that represent the importance of the index terms in the query. [29]

Vector space model has four main methods that will be the core of our study to find the best method among them. These four techniques are

- Inner Product

- Cosine similarity

- Dice Similarity

- Jaccard Similarity

## 1.1.5   IR Applications

IR Applications Information retrieval as one of the NLP models, is used with its techniques in many applications in the recent years. Professional start working on the IR to apply its concepts and techniques to solve the real life problems. [38]

Some of the applications use information retrieval these days are: (1) the Digital Libraries: "Main goals of projects involving digital libraries are the application of the several information retrieval techniques developed in the 80's and the realization of new distributed technology to manage information resources." [42] (2)Media search:

Many websites are using information retrieval to search for multimedia contents, such as: images, videos, news ..etc. (3) Search engines: which are the most popular implementation to the IR system. these search engines have been used by most of internet users. many researches now are trying to transfer from traditional Information Retrieval to Web Information Retrieval. [30]

## 1.2 Outline of the report

The rest of the report is organised as follows :

Chapter 2 : Related Concepts and Literature Review

Chapter 3 : Implementation and Design

Chapter 4 : Experimental Results

Chapter 5 : Conclusion

# Chapter 2

# Literature Review

## 2.1    Related work on Information Retrieval

Working with Information Retrieval Systems and their techniques is quite exciting. Many researches and studies focus on improving the search mechanisms used in IR systems in order to satisfy the user defined query as most as the system can.

The major part of building an IR system is to understand the contents of documents significantly. So "The more the system able to understand the contents of the documents the more effective will be the retrieval outcomes." [41]

One of the most challenge to the researcher in the IR is there ability to build such system that could handle documents written in foreign languages (Non English). This challenge force the specialists to improve the techniques used in IR in order to over come such problem.

Tengku Mohd T. Sembok [41] in his research focused on finding stemming mechanism which improved and increased the effectiveness of the IR system that has been built to work with Arabic and Malay documents. It is common between all languages that the words in general include suffixes, prefixes and infixes. Examples: use , useful, useless , user, etc. The challenge here is to transform both users query and database words into a single unified form, that is known as Conflation . TMT. Sembok try to use Stemming algorithm to conflate morphological variants. By using Conflation process

TMT, Sembok implies that stemming of Malay words has been executed, and found that it performs better than non-conflation method.

Jay Ponte and W.Bruce Croft [23] presented their work on document indexing by building a new approach based on probabilistic language modelling. Guo Xian and et.al [20] proposed a new approach of using information retrieval for online hand written scripts which belong to three scripting families; Arabic, Roman and Tamil scripts. There results come out with accuracy of 93.3%.

Kanaan and et.al [26] suggested a method to improve the Arabic information retrieval using Part of Speech tagging which reduce the indexing storage overhead accordingly. Many Research focused on trying to work on query operation on order to accelerate retrieval process. Gilad Mishne and Maarten de Rijke [35] used in their research phrase and proximity terms for web retrieval rather than traditional ad-hoc retrieval. They implied that the user query entered in web information retrieval has an average with 1.5 to 2.6 terms in length. This went us to the fact that there is a need for an automatic query rewrites, specifically phrasal and proximity-based retrieval, on the performance of web retrieval. Katrin Erk and Sebastian Pad [18] in their exciting research try to highlight the importance of the words meaning in retrieving process.They presented a novel structured vector space model that deals with these issues by choosing favorites for words' argument positions. This could help in integrating the syntax into the computation of word meaning in context.

## 2.1.1    Related work on Vector Space Model

Dr. Khalaf Khatatneh and et.al [15] showed a new mechanism to reduce the space used by the information retrieval system using table memorized semiring structure. In this structure the table includes two arrays one is filled with a word, and the other with some coefficients. This structure implies one document per table, so one row stores the index term and the second row filled with the weight of each index term as illustrated in the below figure 4.2

| D1 | T₁ | T₂ | T₃ | .... | ..... | Tₙ |
|----|----|----|----|------|-------|-----|
|    | W₁₁ | W₁₂ | W₁₃ | ... | ... | W₁ₙ |

Figure 2.1: Table Structure

Using this new mechanism will shrink the gab occurs by using the standard matrix in presenting terms and documents ; The number of rows in the matrix will be added by one when adding new document, and the number of columns will be incremented by one in adding new term as well. But in the table memorized semiring structure, any new document has been added to the text collection means a new separate table will be added to the database, and all new terms will be added to the new table.

Khatatneh and et al implied in their research that using the standard matrix will occupy 204248 units to implement vectors, but the new approach will occupy 5388unit which will save more than 198860 space units.

Michael W. Berry and et al [32] presented in their work how the fundamental mathematical concepts of linear algebra could be used to manage and index such large documents. they assumed that the traditional indexing techniques is useless; since they takes in consideration when indexing any document all the information inside regardless the actual valuable mean for such extracted terms. For example, the abstracts, author lists, titles key word list an so on are some auxiliary information that are not primarily to understand the content of the paper, those are interested in literature search could find such items useful. Therefore, exclude these items from the indexing process will reduce the capacity consuming and improve the retrieval efficiency. Michael W. Berry and et al in their research used the vector space model mathematical equations to implement such approach.

Removing stop words is one of the main phases in the indexing process. Ibrahim Abu El-Khair [16] compared in his research the three stop words lists which are used in Arabic language ( General stoplist, Corpus-based Stoplist and Combined Stoplist). Using these stoplists with Lemur toolkit and multiple weighting schemes, Ibrahim explored the effect of using stop words on Arabic language retrieval. He used different weighting schemes (BM25, KL and TFIDF) and Recall-Precision performance measurement in order to compare the effect the alternative stop words. The result of such comparison study shows that using BM25 weighting schema with combined or general stoplist was the best performing function for Arabic language.

There are many other interesting researches concern about data mining and use vector space models; Grigoreta and Gabriela [36] in their research explained how vector space models could be used in data mining. They presented a new approach that uses clustering in data mining and proposed two techniques: a k-means based clustering technique and a hierarchical agglomerative based clustering technique, and vector space model used for finding the similarity between two methods in order to determine the best over them.

We are in the study focusing on building vector space model that is valid over Arabic language. Working with arabic language to build IR systems is very important; since this will enlarge the arabic contents on the web which is major goal that Arabic researches try to reach.

One of the magnificent researches related to this is done by Ibrahiem El Emary and Ja'far Atwan [17]. Their novel approach was to build such information retrieval that could handle Arabic documents. This comprehensive study used the cosine similarity in the vector space models to compare two technical methods used for retrieving data; these methods are: the full-ward indexing and the root indexing.

The technical way in building their systems done by selecting the number of documents as search data. Then, build the stop word table and finally Build the Inverted table. Their experimental results show that examining the system using the full word indexing method by applying 10 queries against 242 of Arabic texts collection make use of the VSM model with cosine similarity measurement, the proposed system retrieves documents in descending order as shown in table 4.2 :

| Doc. Name | Sim |
|-----------|----------|
| d 47. txt | 0.398656 |
| d 210. txt | 0.294367 |
| d 211. txt | 0.275109 |
| d 177. txt | 0.196763 |
| d 49. txt | 0.154029 |
| d 58. txt | 0.149503 |
| d 55. txt | 0.123651 |
| d 53. txt | 0.07157 |

Figure 2.2: The output of a query search in full word indexing method

While examining the system using root indexing method by applying 10 queries against 242 of Arabic texts collection make use of the VSM model with cosine similarity measurement, the proposed system retrieves documents in descending order as shown in table 4.3 :

To compare between the two methods in term of which is closer to the user need they used precision as a measurement factor to compare the results in both methods and this showed that using root indexing method will give better results than using full word indexing method. As a conclusion Ibrahiem El Emary and Ja'far Atwan proved that that using root indexing is much useful in IR system because of the following reasons: it decreases the size of storage space, minimize the time needed by the system for processing the documents and query, and gives much amount of retrieved data which may satisfy the user query in best manner.

| Doc. Name | Sim |
|-----------|----------|
| d 47. txt | 0.398656 |
| d 210. txt | 0.294367 |
| d 211. txt | 0.275109 |
| d 177. txt | 0.196763 |
| d 49. txt | 0.154029 |
| d 58. txt | 0.149503 |
| d 55. txt | 0.123651 |
| d 53. txt | 0.07157 |

Figure 2.3: The output of query 2 search in root indexing method

# Chapter 3

# Implementation and Design

## 3.1 Vector Space Model techniques and approaches

In this chapter we will explain the four main techniques of vector space model. We will compare between these techniques in retrieving Arabic documents. As mentioned in the introduction there are four main techniques of vector space model :

- Inner Product

- Cosine similarity

- Dice Similarity

- Jaccard Similarity

Inner Product

Inner product is the first technique in the vector space model. This technique considered as a base for other techniques. all the other techniques depend on the results of this technique to compute the results of their functions.

The term inner product space in mathematics is a victor space that has supplementary structure called an inner product.This structure links each couple of vectors in the space with a scalar quantity known as the inner product of the vectors.Inner products let the introduction of intuitive geometrical design for example the length of a vector or the angle between two vectors. Inner product also known as the scalar product. [1] In a vector space, inner product is a way to multiply vectors together. The multiplication of such vectors forms the scalar between them. [2] Figure 3.1 shows the interpretation of the angle between two vectors defined using an inner product.

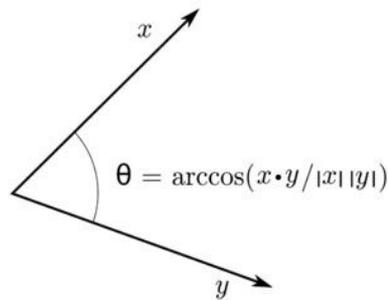

Figure 3.1: interpretation of the angle between two vectors defined using an inner product

The idea here is to implement each document in the retrieval system plus the query as vectors and to find the similarity between these documents and the query. The vector of the documents which are similar in direction to the vector of the query will be considered to be relatives to the query and hopefully satisfy the user need.

Cosine Similarity

Cosine is used as a measure to find the similarity between two vectors. This is done by finding the cosine of the angel between them. [4] So, if the inner product is used to find the distance between two vectors, the cosine is used to find the angel between these vectors.

Using cosine similarity is good way to rank document by finding which document is closer to the user query. Figure 3.2 illustrate the function used to compute the angel between the vectors (documents and query)

$$\text{Sim}(A, B) = \text{cosine } \theta = \frac{A \bullet B}{|A||B|} = \frac{x1'x2 + y1'y2}{(x1^2 + y1^2)^{1/2} \ (x2^2 + y2^2)^{1/2}}$$

Figure 3.2: Compute the Cosine Angel

For example, let say that point A(x1, y1) represents a query and points B(x2, y2), D(x3, y3), E(x4, y4), F(x5, y5), etc represent documents.. We should be able to compute the cosine angle between A (the query) and each document and sort these in decreasing order of cosine angles (cosine similarites). This action can be extended to the whole collection of documents. [6]

Jaccard Similarity

"Also known as Jaccard index, the Jaccard similarity coefficient is a statistical measure of similarity between sample sets" [11] Figure 3.3 shows the function that is used to compute Jaccard similarity [5]

$$s^{(J)}(\mathbf{x}_a, \mathbf{x}_b) = \frac{\mathbf{x}_a^\dagger \mathbf{x}_b}{\|\mathbf{x}_a\|_2^2 + \|\mathbf{x}_b\|_2^2 - \mathbf{x}_a^\dagger \mathbf{x}_b}$$

Figure 3.3: Compute the Jaccard Similarity

Dice Similarity

Dice measurement is used like Jaccard to find the similarity between twe vectors but

"gives twice the weight to agreements" [7] . the measure is given by the formula: [3]

$$D(A,B) = 2|AandB|/(|A| + |B|)$$

where A and B are the two sets. More simply, this formula represents the size of the union of 2 sets divided by the average size of the two sets.

### 3.1.1   Term weighting

All of the above techniques in the vector space model (VSM) have to compute some value called "Term Weight", this expression is the Crucial part in order to retrieve documents and to rank them based on the relevancy of the document to the used need.

There are some main values we have to compute in order to get the term weight.

Term Frequency (TF)

Term Frequency is the standard notion of frequency in natural language processing (NLP); it used to compute the number of times that a term appears in a document. [49] TF expresses the importance of the value in the document and it is widely used since it can be easily calculated inside each document and works well. [37]

Document Frequency (DF)

As the TF represents the importance of the term in the document, the document frequency (df) represent the importance of the term in all documents, it counts the number of documents the term appears in. Document frequency is now used in multiple applications in Natural Language Processing fields such as Information Retrieval and other related fields. [25]

Inverse Document Frequency (IDF)

The Inverse Document Frequency which is considered to be a discriminating measure for a term in the text collection. It was proposed in 1972, and has since been widely used. [40] IDF in information retrieval is used to distinguish words that have the same frequency. [14] IDF is defined mathematically as $idf = \lg_{10}(N/ni), where$

N = the number of documents in the text collection. ni = is the document frequency, the number of documents that contain term i

Tf-idf weighting

We now combine the definitions of term frequency (the importance of each index term in the document(tf) ) and inverse document frequency(the importance of the index term in the text collection), to produce a composite weight for each term in each document.

[12] $w_{ij} = tf_{ij} * idf_i = log_{10}(N/n_i)$

This also called tf-idf scheme, it assigns to word t a weight in document d that is:

- Highest when t occurs many times within a tiny number of documents

- Lower when the term occurs in smaller number of times in a document, or occurs in many documents.

- Lowest when the term exists all documents.

So, consider a text collection consists of 50 documents, a word which appears in each of the documents in the text collection is completely useless term; Since it does not tell any thing about which documents are the user interested in.

### 3.1.2 Performance Measurement

How could we measure the performance of the information retrieval system? This question lead to create some measurements in order to evaluate the IR system. Precision and Recall are used mainly in order to evaluate NLP systems. [22]

In information retrieval system works by running a group of queries against a set of documents and the results retrieved are evaluated using precision, recall and other methods. [47] These evaluation measurements depend on the user decision in order to output there results.

Precision, Recall and others

Effectiveness is the word which is highlighted in term of measuring the ability of the system to satisfy the user need. It is useful at this position to introduce a table which illustrate the relation between the relative and non- relatives documents against the retrieved and not retrieved documents. [45] Figure 3.4 shows such relation.

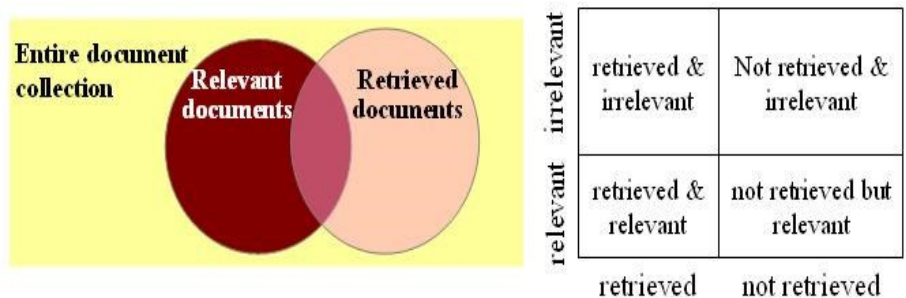

Figure 3.4: Relation Between the relevant and retrieved documents

We can derive from the figure above the equations for both precision and recall as following:

Recall = Number of relevant documents retrieved / Total number of relevant documents Precision = Number of relevant documents retrieved / Total number of docuemnt re-

trieved

Thus, Precision and Recall could be defined as following :

Precision: evaluates the capability of the IR system to retrieve top-ranked documents that are most relevant to the user need, and it is defined to be the percentage of the retrieved documents that are truly relevant to the users query

Recall, on the other hand, evaluates capability of the IR system to get all the relevant documents in the database and is defined as the percentage of the documents that are relevant to the user need [27]

Here in our study we will use precision evaluation measurement in the comparative study in order to evaluate the all techniques since its capable to be tested and to be measured, but it is hard to use the recall in order to evaluate the system since the user has to know all the texts collection in order to compute the evaluation value.

## 3.2    Build Vector Space Model VSM

IR process, as briefly described in the introduction, it is a comprehensive process starts from uploading documents to the system and go through the user need and ends up with retrieving relative information that satisfy the user need. This section will discuss this process in details and shows how this process used to build an IR system using vector space model techniques.

### 3.2.1    Indexing and computing term weight

The following are the steps of indexing the uploaded documents as well as computing the weight for each index term.

Selecting Sample Documents we will use to show the process of building the vector space model an example from page 71 of Grossman and Frieders. [19] Consider the following documents and query:

D1:″‡K  Qmì'@ ú ¯HPQå″  I.ëY   Ë@  á    Ó    é⌡   m…″

D2: ″é'      ®     Ë@  é⌡   kAf  á

Ó   I ÒÊ⌡f@   é'    ®    Ë@  á

Ó   é«A' . ″ D3: ″é⌡   kAf  á   Ó

I ÒÊ⌡f@  I.   ëY  Ë@  á   Ó

é⌡   m…″

Q: ″ é'®Ë@ð I.ëYË@    é⌡    kAf″

Remove Stop-Words

The first step here in order to build the index file is to remove the stop words. Removing stop words and identifying key words are opposite in the goal but relative in the importance of each . [50] Here in our example we will remove those words that considered to be stop words such as "ﻮﻫ , ﻲﻟﺇ@ , ﻥﺃ ﻮﻫ", and will keep the other words of the documents to be index terms used to compute the weight.

Compute the Term Frequency (TF)

Here we will identify the number of times every term appears in a document; i.e the tf for each word. Table 3.1 shows the frequency for all terms in the text collection

Table 3.1: TF

| Document ID | Tf |
|---|---|
| D1 | |
| ﺔﻴﻤ … | 1 |
| ﻝ.ëYË@ | 1 |
| | 1 |
| | 1 |
| HPQﺎ | |
| " | |
| ‡K Qmﻲ'@ | |
| D2 | |
| ﺔ«A | 1 |
| , | 2 |
| | 1 |
| | 1 |
| . ﺔ' | |
| ®Ë@ | |
| ﻝﺅÊ Jf@ | |
| ﺔﻴkA | |
| ﻑ | |
| D3 | |
| ﺔﻴﻤ … | 1 |
| I.ëYË@Jƒ@ | 1 |
| IÒÊ | 1 |
| | 1 |
| éJkA | |
| ƒ | |

Compute the Document Frequency (DF)

Now, we have to compute the number of time the term appears in the text collection which is the (df), as well as the value of the inverse document frequency (idf). as shown in table 3.2 $idf = log_{10}(N/ni)$, where

N = the number of documents in the text collection.

ni = is the document frequency, the number of documents that contain term i

Table 3.2: Inverse Document Frequency

| Term ID | Term | DF(ni) | Idf=N/idf |
|---|---|---|---|
| t1 | IÒÊJƒ@ | 2 | log10 3/2 = 0.176 |
| t2 | HPQå" | 1 | 0.477 |
| t3 | é«A ' . | 1 | 0.477 |
| t4 | ‡K Qmì'@ | 1 | 0.477 |
| t5 | I.ëYË@ | 2 | 0.176 |
| t6 | é'®Ë@ | 1 | 0.477 |
| t7 | éJm ... | 2 | 0.176 |
| t8 | éJkA ƒ | 2 | 0.176 |

Compute the weight

After that, we will compute the weight for each term as shown in table 3.3

$Wij = tf_{ij} * idf_i = tf_{ij} * log_{10}(N/ni)$

Table 3.3: The weight for each term

| | T1 | T2 | T3 | T4 | T5 | T6 | T7 | T8 |
|---|---|---|---|---|---|---|---|---|
| D1 | 0 | 1*0.477 = 0.477 | 0 | 1*0.477 = 0.477 | 1*0.176 = 0.176 | 0 | 1*0.176 = 0.176 | 0 |
| D2 | 0.176 | 0 | 0.477 | 0 | 0 | 2*0.477 = 0.945 | 0 | 0.176 |
| D3 | 0.176 | 0 | 0 | 0 | 0.176 | 0 | 0.176 | 0.176 |
| Q | 0 | 0 | 0 | 0 | 0.176 | 0.477 | 0 | 0.176 |

### 3.2.2    Apply vector space model techniques

After we have built our index file and compute the weight for all the index terms in our text collection, we will apply the four vector space model techniques in order to find the degree of similarity between the query and every documents in order to rank the retrieved documents as for each technique.

Inner Product (Dot Product) Similarity

In dot product similarity we will compute the similarity between vectors for the document di and the query q based on the following equation:

$$\sum_{k=1}^{t}(d_{ik}.q_{k})$$

where $d_{ik}$ is the weight of term i in the document k and $q_k$ is the weight of term i in the query.

Now we will apply the above equations in order to compute the degree of similarity between each document and the query based on dot product technique:

$SC(Q,D1)$ = (0)(0)+(0)(0.477)+(0)(0)+(0)(0.477)+(0.176)(0.176)+(0)(0.477)+ (0)(0.176) + (0)(0.176) = 0.031

$SC(Q,D2)$ = 0.486

$SC(Q,D3)$ = 0.062

Now we will rank the retrieved documents in a descending order and use threshold to retrieve documents above the value of that threshold.

Cosine Similarity

As we introduced the cosine similarity, it is used to compute the angel between the vectors for documents and query. We will apply the below equation of cosine similarity on our example:

$$\frac{\sum_{k=1}^{t}(d_{ik} \cdot q_k)}{\sqrt{\sum_{k=1}^{t}(d_{ik})^2 * \sum_{k=1}^{t}(q_k)^2}}$$

- Where dik is the weight of term i in document k and qk is the weight of term i in the query.

- $\sum_{k=1}^{t}(d_{ik} \cdot q_k)$ is the inner product and

- $\sqrt{\sum_{k=1}^{t}(d_{ik})^2 * \sum_{k=1}^{t}(q_k)^2}$ is the length for documents and query.

Note: When using cosine similarity measure, computing the tf-idf values for the query terms we divide the frequency by the maximum frequency(2) and multiply with the idf values. So the query will become as in table 3.4

Table 3.4: The new weight for each term in the query

| | T1 | T2 | T3 | T4 | T5 | T6 | T7 | T8 |
|---|----|----|----|----|----|----|----|----|
| Q | 0 | 0 | 0 | 0 | $\frac{1}{2}$*0.176= 0.088 | $\frac{2}{2}$*0.477 = 0.477 | 0 | $\frac{1}{2}$*0.176 = 0.088 |

Now we will compute the length for each documents :

Length of D1 = $\sqrt{0.477^2 + 0.477^2 + 0.176^2 + 0.176^2}$ = 0.7195

Length of D2 = $\sqrt{0.176^2 + 0.477^2 + 0.954^2 + 0.176^2}$ = 1.095

Length of D3 = $\sqrt{0.176^2 + 0.176^2 + 0.176^2 + 0.176^2}$ = 0.352

Length of Q =    $0.088^2 + 0.477^2 + 0.088^2 = 0.495$

Inner product for each document is:

$D1 = 0.031$     $D2 = 0.486$     $D3 = 0.062$

*Then the similarity values will be :

$cosSim(D1, Q) = \frac{0.0309}{0.719*0.495} = 0.087$

$cosSim(D2, Q) = \frac{0.485}{1.095*0.495} = 0.9$

$cosSim(D3, Q) = \frac{0.061}{0.352*0.495} = 0.357$

Finally, sorting the above results in descending order will give document (2) the top ranked since it has the maximum similarity value with the query, then document (3) , then document (1) which will get the lowest rank since it has the lowest similarity value with the query.

Jaccard Similarity

We will continue with our example to find the similarity values between the documents ans the vectors using Jaccard Similarity as the following formula :

$$\frac{\sum_{k=1}^{t}(d_{ik}.q_k)}{\sum_{k=1}^{t}(d_{ik})^2 + \sum_{k=1}^{t}(q_k)^2 - \sum_{k=1}^{t}(d_{ik}.q_k)}$$

Where $\sum_{k=1}^{t}(d_{ik}.q_k)$ is the inner product and $\sum_{k=1}^{t}(d_{ik})^2$ is the square of summation of weight for each term in each document and $\sum_{k=1}^{t}(q_k)^2$ which is the summation of weight of each term in the query.

* The square of the summation of the weight of each term in each document is :

- $\sum_{k=1}^{8}(d_{1,k})^2 = (0.477^2 + 0.477^2 + 0.176^2 + 0.176^2) = 0.517$

- $\sum_{k=1}^{8}(d_{2,k})^2 = (0.176^2 + 0.477^2 + 0.954^2 + 0.176^2) = 1.2$

- $\sum_{k=1}^{8}(d_{3,k})^2 = (0.176^2 + 0.176^2 + 0.176^2 + 0.176^2) = 0.124$

* Square the summation of the weight of each term in the query is :

- $\sum_{k=1}^{8}(q_k)^2 = (0.176^2 + 0.477^2 + 0.176^2) = 0.289$

$\sqrt{}$

Then the similarity values based on Jaccard are:

- $JacSim(D1, Q) = \frac{0.0309}{0.517 + 0.289 - 0.0309} = 0.04$

- $JacSim(D2, Q) = \frac{0.486}{1.2 + 0.289 - 0.486} = 0.485$

- $JacSim(D3, Q) = \frac{0.062}{0.124 + 0.289 - 0.062} = 0.177$

** After that, we rank the documents in descending order according to the similarity score, as shown in table 3.5

Table 3.5: Jaccard Similarity

| Document ID | Similarity Score |
| --- | --- |
| D2 | 0.485(most relevant) |
| D3 | 0.177 |
| D1 | 0.04 |

Dice Similarity

The last technique which we are going to implement is the Dice similarity. The following is the main formula for it:

$$\frac{2 * \sum_{k=1}^{t}(d_{ik} \cdot q_k)}{\sum_{k=1}^{t}(d_{ik})^2 + \sum_{k=1}^{t}(q_k)^2}$$

Where $\sum_{k=1}^{t}(d_{ik} \cdot q_k)$ is the inner product and $\sum_{k=1}^{t}(d_{ik})^2$ is the square of summa-tion of weight for each term in each document and $\sum_{k=1}^{t}(q_k)^2$ which is the summation of weight of each term in the query.

\* The square of the summation of the weight of each term in each document is :

- $\text{P}_{8_{k=1}}(d_{1,k})^2 = (0.477^2 + 0.477^2 + 0.176^2 + 0.176^2) = 0.517$

- $\text{P}_{8_{k=1}}(d_{2,k})^2 = (0.176^2 + 0.477^2 + 0.954^2 + 0.176^2) = 1.987$

- $\text{P}_{8_{k=1}}(d_{3,k})^2 = (0.176^2 + 0.176^2 + 0.176^2 + 0.176^2) = 0.124$

\* Square the summation of the weight of each term in the query is :

- $\sum_{k=1}^{8}(q_k)^2 = (0.176^2 + 0.477^2 + 0.176^2) = 0.289$

$\sqrt{}$

Then the similarity values based on Dice similarity are:

$DiceSim(D1, Q) = \frac{2*0.0309}{0.517+0.289} = 0.077$

- $DiceSim(D2, Q) = \frac{2*0.486}{1.2+0.289} = 0.653$

- $DiceSim(D3, Q) = \frac{2*0.061}{0.124+0.289} = 0.3$

\*After that, we rank the documents in descending order according to the similarity score, as shown in table 3.6

Table 3.6: Dice Similarity

| Document ID | Similarity Score |
| --- | --- |
| D2 | 0.653(most relevant) |
| D3 | 0.3 |
| D1 | 0.077 |

Retrieval Performance Evaluation

As we stated regarding the performance measures; Precision and Recall are used to evaluate information retrieval system. Figure 3.5 illustrate the relation between the retrieved and the relevant documents

So, the precision and recall could be computed based on the following :

*Recall = Numberofrelevantdocumentsretrieved/Totalnumberofrelevantdocuments*

*Precision = Numberofrelevantdocumentsretrieved/Totalnumberofdocuemntretrieved* For our example, Assume that the number of relevant document which retrieved is 1 , total number of relevant document is 2, total number of document retrieved is 3.

*Recall = 1/2 = 0.5*

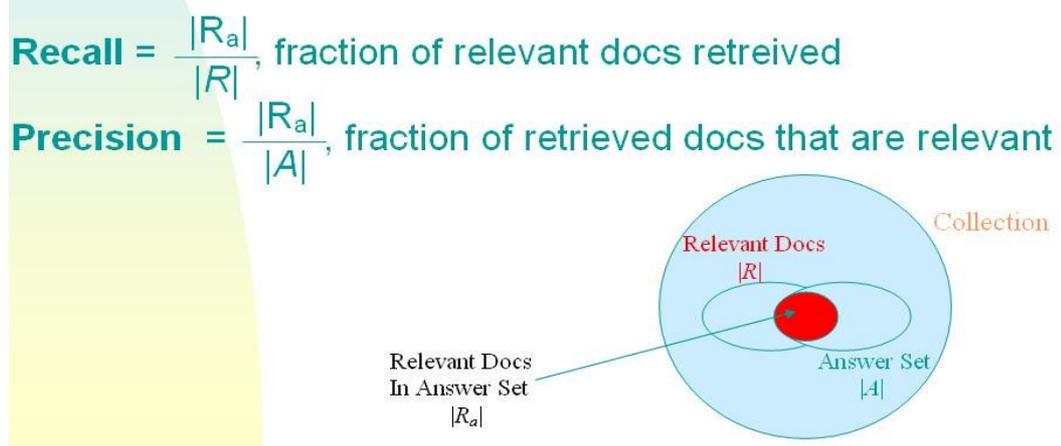

Figure 3.5: Relation Between the relevant and retrieved documents

*Precision = 1/3 = 0.333*

We have represented and explained in this chapter the main steps for implementing an Information retrieval system. We will show on the next chapter the web tool used to apply the four techniques in vector space model and the experimental results as well.

# Chapter 4

# Experimental Results

## 4.1   Why web-based tool?

Cloud computing is now a hot topic that is discussed widely in many conferences. clouding where applications and documents are hosted on a cloud that consists of thousands of computers and servers machines, all linked to be accessible via the internet. Using the clouding technology you can now access every single file through the internet instead of being a desktop based. [34]

For this reason we built our tool using web environment rather than desktop environment. The core programming language that was used in the project is PHP. which stands for Hypertext Preprocessor, is mainly web purpose scripting languages that create dynamic web sites. [46]. One of its main features that it is an open source language and it is secure language as well.More over, PHP could run on (almost) any platform and since PHP uses the Apache server this makes PHP very fast. [8]

Since our tool is a web-based, we used other web programming languages and scripting such as html, javascript and cascading style sheet (css). In order to create a dynamic database website, Our information retrieval system is connected to MySql database management system in order to store all the terms, term-frequencies , term-weights and the query information as well.

## 4.2    IR-VSM tool

We will show in this section snapshots of the main parts of our website. and the process of building the information retrieval system in GUI in order to make the whole process clear.

Index Page

Figure 4.1 shows the home page of our web site.

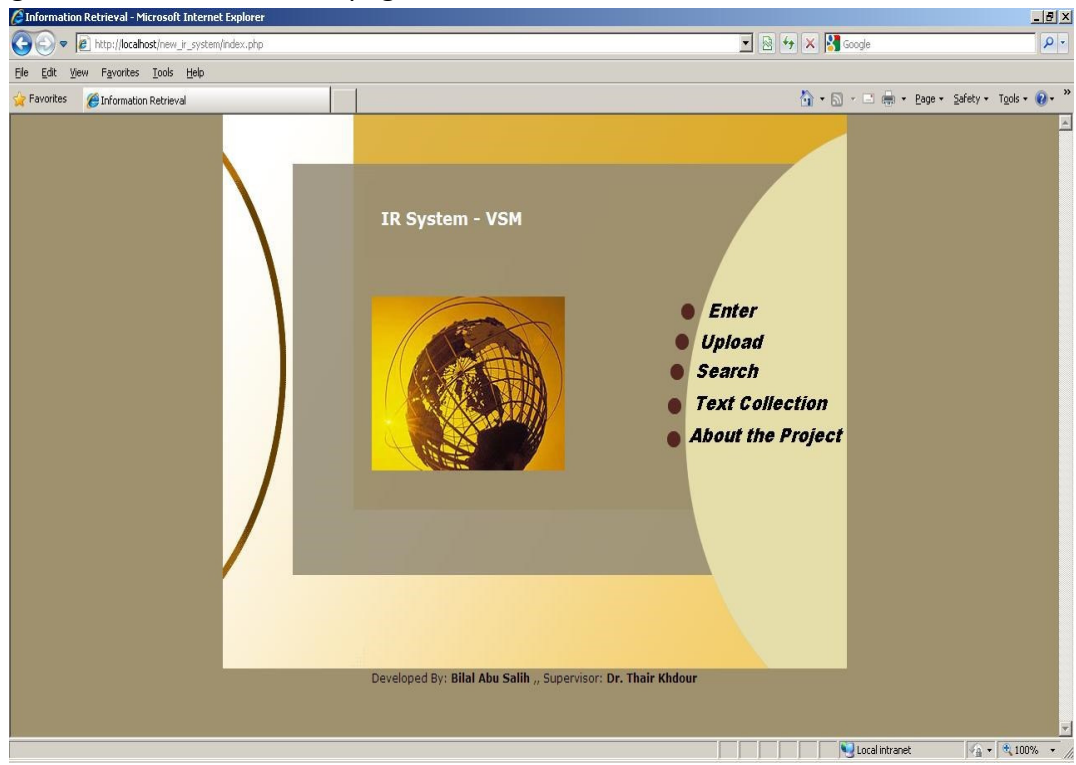

Figure 4.1: Index page

Here the user will chose one of the following items in order to start browsing our tool :

- Enter : This will take the user to the home page of our website

- Upload: Will take the user to the upload documents page.

- Search: Where the user can search for relative documents in the repository

- Text Collection: Here the user can browse the uploaded documents instead of searching for specific document.

- About the project:This Gives a brief about this project and some acknowledgements.

Home

Figure 4.2 hows the home page where the user can read briefly about the vector space model in information retrieval system and gives some hints before starting.

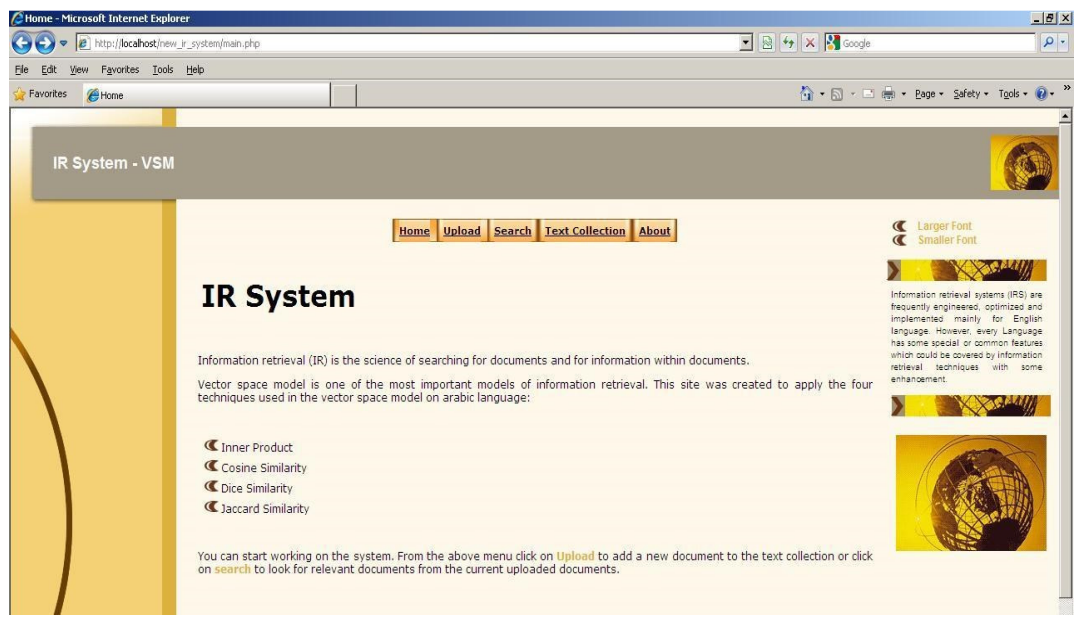

Figure 4.2: Home page

Upload

Figure 4.3 shows the upload page which includes some text on the right side explained the process of indexing and the importance of such phase of Information retrieval.

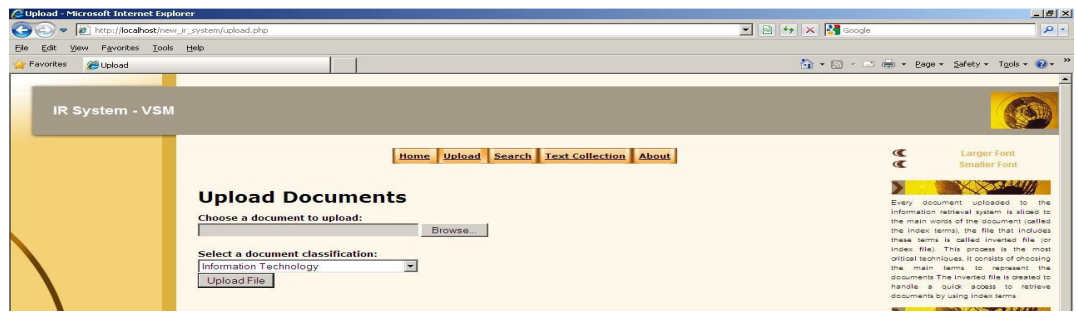

Figure 4.3: Upload Page

In this page you feed the repository with documents by selecting them from browse button. then chose what classification such document belongs to as shown in figures

4.4 and 4.5

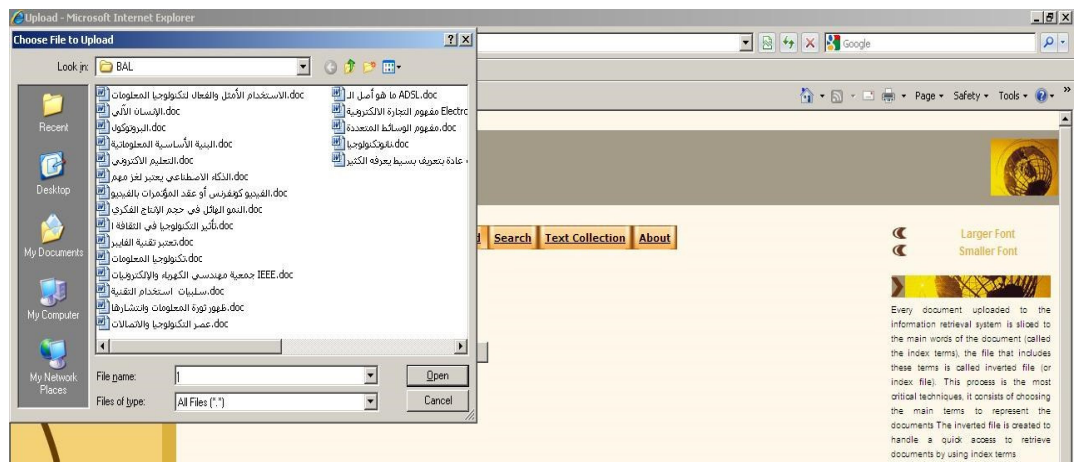

Figure 4.4: Browse

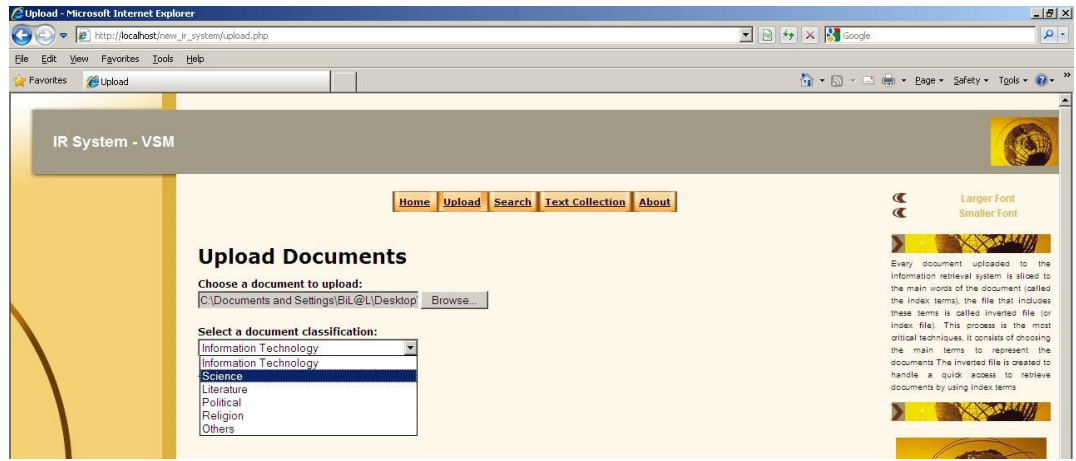

Figure 4.5: Select Classification

After the document is selected, it will go via an inverter code that will extract from the document the main terms and store them in a database table.

Once the document is uploaded successfully the following message appears.

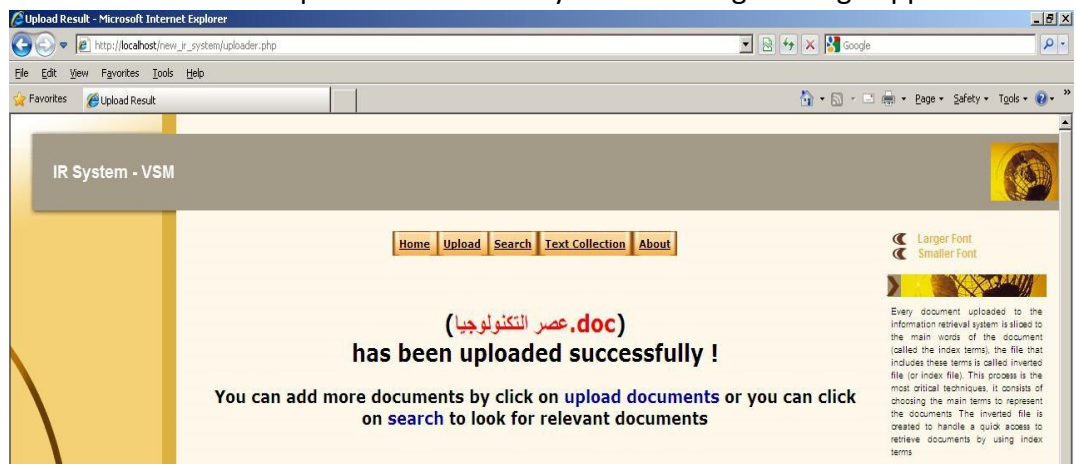

Figure 4.6: Upload Result page

Search

To search for any relative document, you have to click in the search button form the above menu, this will open the page illustrated in figure 4.7.

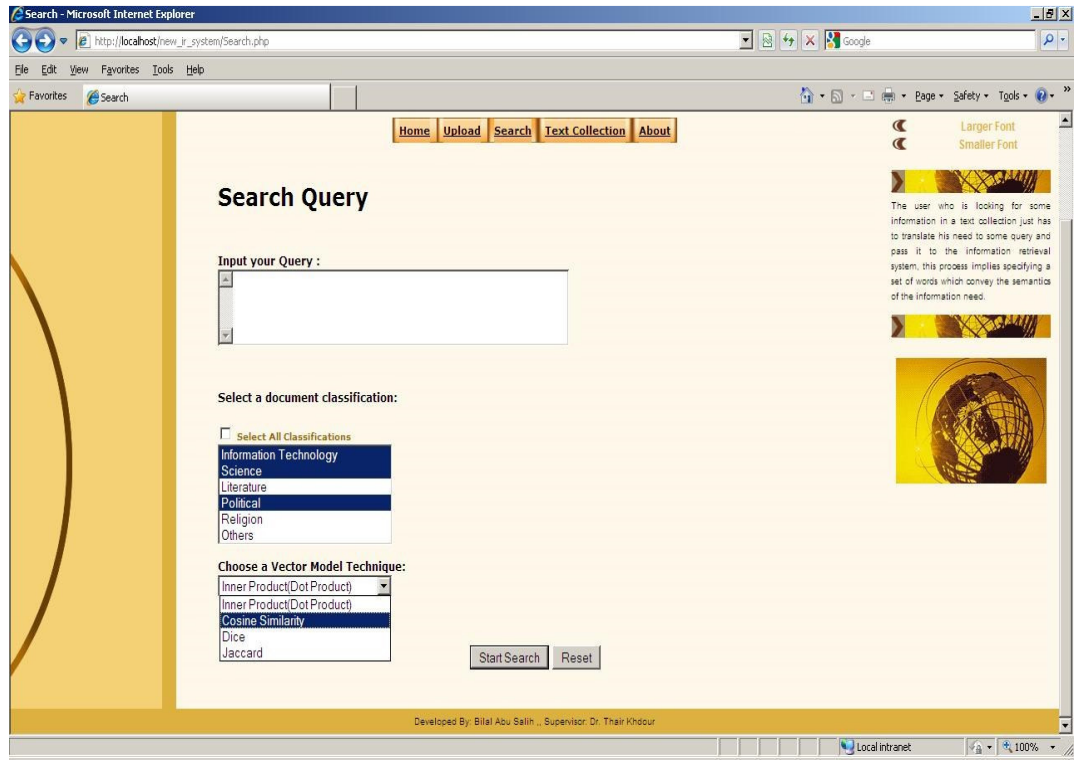

Figure 4.7: Search page

The main core of this website is the ability to search for the user query based on one of the vector space models techniques.

The user will chose from the list of classifications one or some classifications and chose one of vector space model techniques then click "Start Search". The search criteria will be passed via certain codes to look for the documents that are relevant to the query entered. The search results for the user's query as per the search criteria will be as shown in figure 4.9

Figure 4.8: Search Results page

    We can figure from the above snapshot that the retrieved documents are sorted out as per their ranking value. One of the important features in this page is the ability to judge the retrieved document, as per the relevancy of the document to the user need. So, once the user check the check box in the right of each documents in the search results, this will rise the precision value and get feed back that the document is relevant.

Text Collection

We can browse the new uploaded document or the current uploaded documents by clicking on the text collection from the above menu. The text collection will look like the below snapshot:

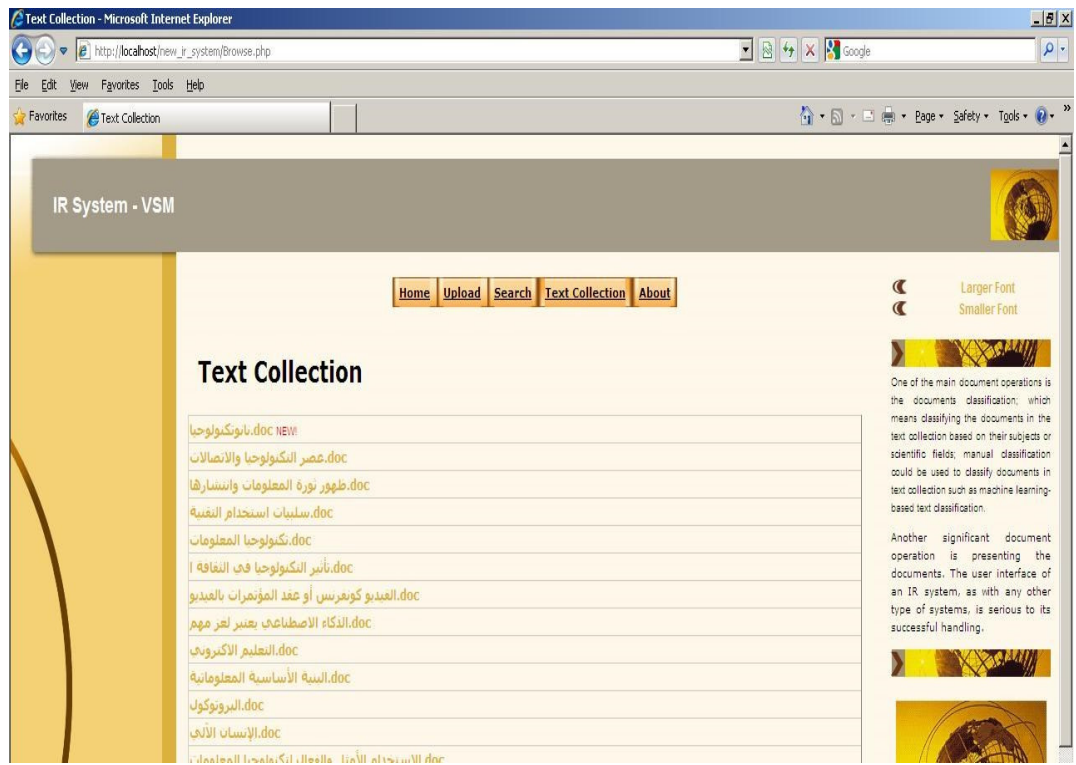

Figure 4.9: Text Collection page

We have presented in this chapter the information retrieval web tool and the main features of it. Using this web tool is magnificent and interesting; since the user could do the following:

- Upload Arabic documents to the repository of the web site

- Search for relative documents and give a relevance feedback.

- Browse the text collection

- And Even more

# Chapter 5

# Conclusion

We have demonstrated in this report an implementation of Information Retrieval system. Working with such scientific field is motivating and significant; since many researches try to develop information retrieval systems that could handle documents written in specific language.

This paper is significant in many terms, we have presented the main features of Vector Space Model and we implemented these features using web technology to cover documents written in Arabic languages.

We will work in the future to improve the tool in order to enlarge its features, such as covering pdf, html and other file format. Also, give the system the ability to detect the type of device from which the request has been sent in order to handle the request in effective manner and give the user the ability to browse this web tool based on the capabilities of such device. Moreover, we have to make our website secure by limiting access for some features in the website to the anonymous user and allow these features to the granted users only.

We are glad to build such tool and we do appreciate all the readers of this report to try out this tool and send their feedback in order to enhance the current tool and help out Arabic users to significantly use this tool in efficient manner.